\documentclass[12pt]{article}
\usepackage{times}
\usepackage{geometry}
\usepackage[utf8]{inputenc}
\usepackage{enumitem,amssymb}
\usepackage{ragged2e}
\newlist{thematic}{itemize}{8}
\setlist[thematic]{label=$\square$}
\usepackage{pifont}
\usepackage{graphicx}
\usepackage{natbib}
\usepackage{aas_macros}
\usepackage{titlesec}

\titleformat{\section}
  {\normalfont\bf}{\thesection}{1em}{}
  
\begin{document}
\raggedright
\pagenumbering{roman}
\huge
Astro2020 Science White Paper \linebreak

Protoplanetary Disk Science Enabled by Extremely Large Telescopes
\linebreak
\normalsize

\noindent \textbf{Thematic Areas:} \hspace*{60pt} $\boxtimes$ Planetary Systems \hspace*{10pt} $\boxtimes$ Star and Planet Formation \hspace*{20pt}\linebreak
$\square$ Formation and Evolution of Compact Objects \hspace*{31pt} $\square$ Cosmology and Fundamental Physics \linebreak
  $\square$  Stars and Stellar Evolution \hspace*{1pt} $\square$ Resolved Stellar Populations and their Environments \hspace*{40pt} \linebreak
  $\square$    Galaxy Evolution   \hspace*{45pt} $\square$             Multi-Messenger Astronomy and Astrophysics \hspace*{65pt} \linebreak
  
\textbf{Principal Author:}

Name:	Hannah Jang-Condell
 \linebreak						
Institution:  University of Wyoming
 \linebreak
Email: hjangcon@uwyo.edu
 \linebreak
Phone:  307-766-6150
 \linebreak
 
\textbf{Co-authors:} (names and institutions)
  \linebreak
  Sean Brittain (Clemson University), % sbritt@clemson.edu
  Alycia Weinberger (Carnegie Inst of Washington),
  Michael Liu (University of Hawaii), % mliu@ifa.hawaii.edu
  Jacqueline Faherty (American Mus of Nat History), % jfaherty@amnh.org
Jaehan Bae (Carnegie Inst of Washington), % jbae@carnegiescience.edu
Sean Andrews (Harvard-Smithsonian Center for Astrophysics), % sandrews@cfa.harvard.edu
Megan Ansdell (UC Berkeley), % ansdell@berkeley.edu
Til Birnstiel (LMU Munich), % til.birnstiel@lmu.de
Alan Boss (Carnegie Inst of Washington), % aboss@carnegiescience.edu
Laird Close (University of Arizona), % lclose@as.arizona.edu
Thayne Currie (NASA-Ames / Subaru Telescope), % currie@naoj.org
Steven J Desch (Arizona State University), % steve.desch@asu.edu
Sarah Dodson-Robinson (University of Delaware), % sdr@udel.edu
Chuanfei Dong (Princeton University), % dcfy@princeton.edu
Gaspard Duchene (University of California Berkeley), % gduchene@berkeley.edu
Catherine Espaillat (Boston University), % cce@bu.edu
Kate Follette (Amherst College), % kfollette@amherst.edu
Eric Gaidos (University of Hawaii), % gaidos@hawaii.edu
Peter Gao (University of California Berkeley ), %gaopeter@berkeley.edu
Nader Haghighipour (Institute for Astronomy, University of Hawaii), % nader@ifa.hawaii.edu
Hilairy Hartnett (Arizona State University), % h.hartnett@asu.edu
Yasuhiro Hasegawa (Jet Propulsion Laboratory), % yasuhiro.hasegawa@jpl.nasa.gov
Mihkel Kama (University of Cambridge), % mkama@ast.cam.ac.uk
Jinyoung Serena Kim (University of Arizona), % serena@as.arizona.edu
\'{A}gnes K\'{o}sp\'{a}l (Konkoly Observatory, Budapest, Hungary), % kospal@konkoly.hu
Carey Lisse (Johns Hopkins University Applied Physics Laboratory), % carey.lisse@jhuapl.edu
Wladimir Lyra (California State University, Northridge), % wlyra@csun.edu
Bruce Macintosh (Stanford University), % bmacint@stanford.edu
Dimitri Mawet (California Institute of Technology), % dmawet@astro.caltech.edu 
Peregrine McGehee (College of the Canyons), % peregrine.mcgehee@canyons.edu
Michael Meyer (University of Michigan), % mrmeyer@umich.edu
Eliad Peretz (NASA Godard Flight Center), % eliad.peretz@nasa.gov
Laura Perez (Universidad de Chile), % lperez@das.uchile.cl
Klaus Pontoppidan (Space Telescope Science Instutute), % pontoppi@stsci.edu
Steph Sallum (University of California at Santa Cruz),% ssallum@ucsc.edu
Colette Salyk (Vassar College), % cosalyk@vassar.edu
Andrew Szentgyorgyi (Harvard-Smithsonian Center for Astro.), % saint@cfa.harvard.edu
Kevin Wagner (University of Arizona) % kwagner@as.arizona.edu
\linebreak  

\textbf{Abstract:}

The processes that transform gas and dust in circumstellar disks into
diverse exoplanets remain poorly understood.  One
key pathway is to study exoplanets as they form in their young
($\sim$few~Myr) natal disks.  Extremely Large Telescopes (ELTs)
such as GMT, TMT, or ELT, can be used to establish the initial chemical
conditions, locations, and timescales of planet formation, via
(1)~measuring the physical and chemical
conditions in protoplanetary disks using infrared spectroscopy and
(2)~studying planet-disk
interactions using imaging and spectro-astrometry.
Our current knowledge is based on a limited
sample of targets, representing the brightest, most extreme cases,
and thus almost certainly represents an
incomplete understanding.  ELTs will play a
transformational role in this arena, thanks to the high
spatial and spectral resolution data they will deliver.
We recommend a key science program to conduct a volume-limited survey of 
high-resolution spectroscopy and high-contrast imaging 
of the nearest protoplanetary disks
that would result in an unbiased, holistic picture
of planet formation as it occurs.

%In order to better understand the nature of exoplanets and their wide variety, we propose to %study their origins: that is, the formation of exoplanets in the gas-rich protoplanetary disks %seen around young pre-main sequence stars.  Because these young systems are typically distant %($\gtrsim140$ pc), studying them with current state-of-the-art 8-10 m class telescopes remains %a challenge because of the faintness and small angular size of the disks.  This means that %currently only a few bright targets are studied in much detail, making it difficult to draw %general conclusions about the planet formation process.  ELTs will play a transformational %role in providing increased sensitivity and angular resolution for studying planet-forming %disks. 
\pagebreak

\setlength{\parindent}{2em}
\pagenumbering{arabic}
\vspace{-1ex}
\section{Introduction}
\vspace{-2ex}
The processes that transform gas and dust in circumstellar disks into diverse exoplanets are poorly understood. 
Existing theories of disk chemistry invoke different assumptions (about grain sizes, heating and cooling processes, molecular formation and destruction rates), and few models have been validated observationally to determine whether the assumptions are correct. 
Similarly, imaging observations of disks are not yet of high enough spatial resolution to provide good constraints to models of planet/disk interaction. 
A unified theory of planet formation requires information on conditions within the progenitor disks and studies of newly formed planets at the same scale ($<10$AU) of current exoplanet surveys.  

\begin{figure}[b!]
%\begin{figure}[h!]
    \centering
    \vspace*{-1ex}
    \includegraphics[width=0.95\textwidth]{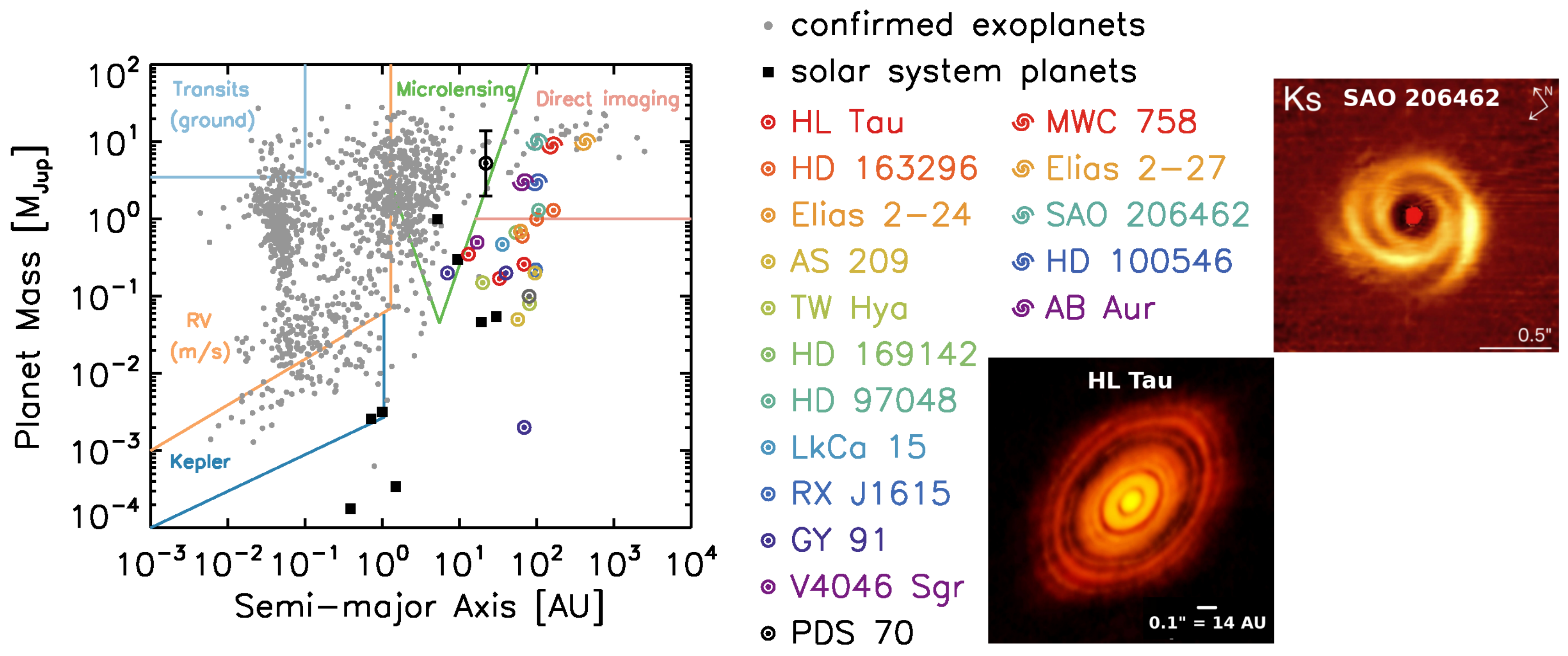}
    \vspace*{-2ex}
    \caption{\label{fig:substructure} \em
    Mass vs.~orbital distance diagram of exoplanets and putative protoplanets in disks with gaps ($\odot$ symbols) or spiral arms (spiral symbols), adopted from \citet{bae18}. The two inset images show example disks with gaps and spirals: HL Tau \citep{alma2015} and SAO~206462 \citep{garufi2013}.
    Putative protoplanets are located outside of the parameter space in which current exoplanet detection techniques are capable of finding exoplanets. With ELTs, we will be able to peer into the innermost regions of planet-forming disks.
    }
\end{figure}

To date, protoplanetary disks have only been probed in detail in their outer regions, whereas exoplanet discoveries are mostly inside those radii (Fig.~\ref{fig:substructure}). 
Transformational studies of planet formation will require studying protoplanetary disks at the very high spatial and spectral resolution provided by 
20-m to 30-m telescopes such as the Giant Magellan Telescope (GMT),  
the Thirty Meter Telescope (TMT), 
and the European Extremely Large Telescope (ELT). 
These Extremely Large Telescopes (ELTs) have both the sensitivity 
(due to the large collecting area) and angular resolution to reveal new 
details about protoplanetary disks at unprecedented scales.  
Such work will establish the chemical initial conditions, locations, and timescales of planet formation.  In this white paper, we focus specifically on the 
science cases to:
%\begin{enumerate}[nolistsep]
%\item 
(1) probe physical and chemical conditions in protoplanetary disks at the scale of planet formation using infrared spectroscopy, and 
%\item 
(2) study planet-disk interactions through imaging and spectro-astrometry.  
%\end{enumerate}

\vspace{-2ex}
\section{Physical and Chemical Conditions of Planet Formation}
\begin{figure}[tb]
\centering
\includegraphics[width=0.7\textwidth]{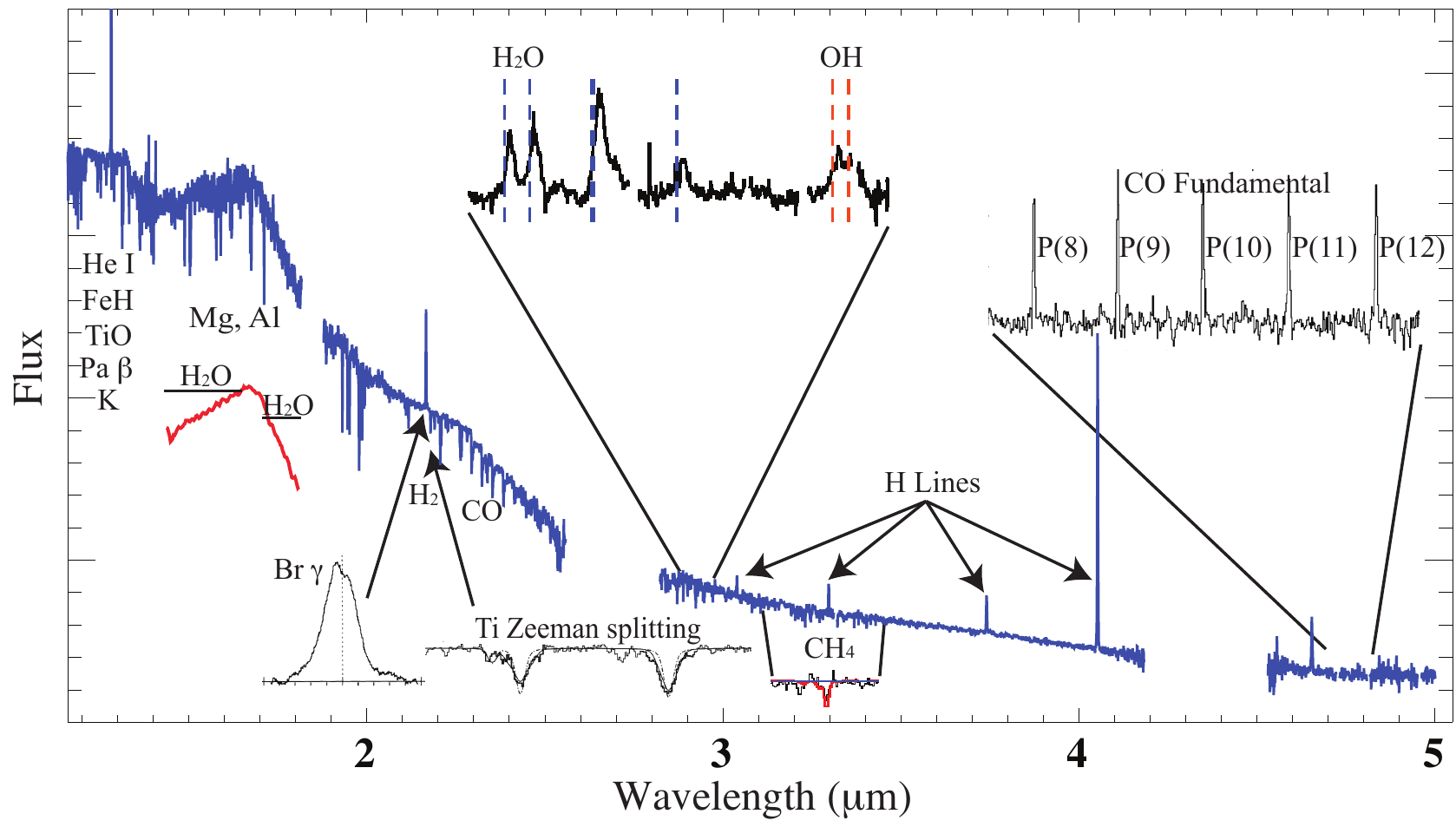}\vspace{-1.5ex}
\caption{\label{fig:spectrum} \em
%Infrared spectroscopy can probe many aspects of the chemistry and dynamics of disks at scales relevant to planet formation given the spatial and spectral resolution of ELTs. The near-infrared data shown demonstrate how the nearest classical T Tauri star, TW Hydrae, can be studied on a disk-averaged basis. 
Near-infrared spectra of TW Hydrae, the nearest classical T Tauri star.
Here, a low resolution spectrum in blue \citep{VaccaSandell:2011} is overlayed with a subset of the interesting spectral features available for young stars, and circumstellar disks. Hydrogen emission line profiles and strengths probe disk accretion physics \citep{Najita:1996}, spectral shape and atomic line widths indicate age \citep{Allers:2013}, Zeeman splitting measures stellar magnetic fields \citep{Yang:2008}, molecular emission lines reveal disk kinematics, abundances, and temperature structure \citep{Banzatti:2017,Gibb:2013,Salyk:2007}. Regions of strong telluric absorption are not plotted. 
The ELTs will enable the more typical young stars at $\sim$140 pc to be studied with the same fidelity and also with spatial resolution. Credit: A. Weinberger.}
    \vspace*{-2ex}
\end{figure}

\begin{figure}[tb]
    \centering
    \includegraphics[width=0.95\textwidth]{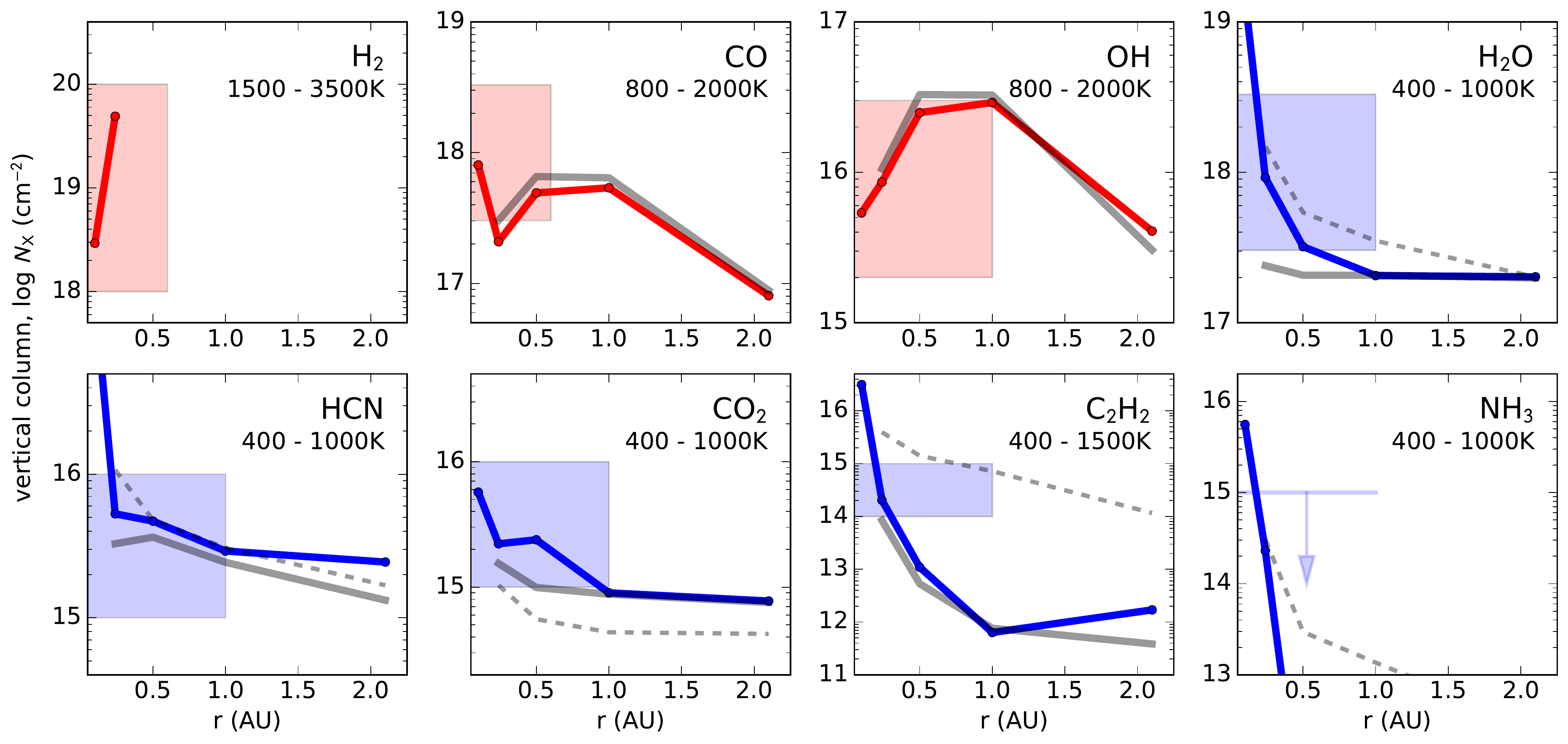}
    \vspace{-2ex}
    \caption{\label{fig:chem_abund} \em
%Most previous work on disks is both spatially and spectrally unresolved. The ELTs will provide spatial information about chemistry in the inner disk. In this figure, the shaded boxes indicate the range of disk-average abundances of primary gas phase species as measured with spectroscopy and the solid lines show simple disk structure models that fit the data (Najita \& Adamkovics 2017). Variation between disks will be important to the variation of compositions of resulting planets. ELTs equipped with high-resolution spectrographs will provide spatial information about the chemistry, probe a wider range of disks around stars of lower mass, and assess the natural variation in chemistry between disks. 
Abundance of the primary molecular gas phase species in disks 
\citep{Najita2017}. The color lines plot  the vertical column density of the species as a function of
stellocentric radius. The gray solid line reflects the abundance in the absence of mechanical heating
in the disk. The gray dashed line reflects the abundance in the absence of Ly-$\alpha$ emission in the
disk. The shaded boxes indicate the measured abundance of each molecule. ELTs will provide
more sensitive measures of the composition of the inner disk for an unbiased sample of sources.
  }  
    \vspace*{-2ex}
\end{figure}

\vspace{-2ex}
Whether a planet is wet or dry, icy or rocky, depends on how and how fast material is processed through a disk. The compositions of the feeding zone for planets that eventually populate the habitable zone of low-mass stars ($<$1 M$_\odot$) span scales of a few tenths to 10 AU. The temperatures of interest will be $\lesssim$500~K, so that refractory solids are condensed throughout the region while volatile condensation fronts exist at the outer edges (low temperatures). Here, disks typically have high gas column densities and optical depths and only the upper disk atmosphere can be observed in the optical/near-IR (see Fig.~\ref{fig:spectrum}). However, given short timescales for vertical mixing, this region should be closely linked to the larger molecule-rich disk interior.  Line widths and velocities for kinematic measurements of circulation (turbulence) and accretion in the disk surface are necessary to make that connection quantitative. JWST will make strides on the warm molecular layer and optically thin holes, but lacks high spectral resolution for determining gas kinematics. ALMA is making outstanding progress on the cool, outer disk and disk midplane. The role of the ELTs will be to provide higher spatial and spectral resolution than JWST such that $<$10 AU scales are reachable for more than just the most extreme disks accessible with existing 8-m class facilities (see Fig.~\ref{fig:chem_abund}).

Carbon-bearing species and water are of particular interest.  
Measurements of the major carbon-carrying molecules (CO, C$_2$H$_2$, CH$_4$) and atomic carbon, simultaneously, will help us understand the carbon depletion of terrestrial planets \citep{Gail2017}, and predict the carbon abundance of terrestrial exoplanets.
Spectroscopy of H$_2$O will allow us to understand the role of water ice in the formation of gas giants, and to measure the profile of C/O in the natal disk.  Mid-IR spectra can provide the abundance of H$_2$O within the so-called ``snow line'', and, with sufficient coverage at long wavelengths, also provide the snow line location \citep[e.g.][]{Zhang2013,Blevins2016}. Observations of OH can also provide insight into the chemical formation and destruction of water vapor in terrestrial planet forming regions, including measuring how much ultraviolet is penetrating the disk's upper layers to drive chemistry \citep[e.g.][]{Najita2010}.  
The increased sensitivity provided by ELTs will make it possible to study the detailed distribution of gas species
%, and therefore the important C/O ratio, 
with distance from the star for low-mass host stars in nearby ($<$200 pc) star-forming regions. 
The C/O ratio sets the oxidation state of a planet and may change drastically based on where a planet or its constituents form \citep[e.g.][]{Oberg2011}. 
Derived temperatures and gas column densities will test models of disk thermal structure and chemistry, revealing the chemical and physical initial conditions for planet formation.

Few disks currently have direct measures of their turbulence, but this is a fundamental property of disks that influences the growth and settling of grains, the vertical thermal profiles and chemical mixing between the midplane and surface,  and rate of accretion of grains and pebbles into the star.  The relative opacities of closely spaced optically thick emission lines can reveal the ``microscopic" turbulence of the gas. 
%% If gas temperature can be inferred from widely-spaced lines, then the excess micro-motion can be distinguished from the macro-bulk motion of the gas. 
This method has been applied to infer the turbulence at the inner rim of disks \citep{Hartmann2004,Carr2004}. 
ALMA observations of {\it outer} disks have found surprisingly low levels of turbulence, and it has yet to be fully understood how these can be consistent with observed rates of stellar accretion \citep[e.g.][]{Hughes2011,Flaherty2017}.  
With ELTs and very high spectral resolution, it may also be possible to use the abundant 3$\mu$m water lines in the L-band. 
The sensitivity and large spectral grasp of high resolution spectrographs on ELTs will enable the study of {\it inner} disk turbulence that will bridge the gap between what has been learned about the inner rim of disks and the outer disk.

\vspace{-2ex}
\section{Planet-disk Interactions}
\vspace{-2ex}
Substantial feedback occurs between forming planets and disks that affects disk structure the incorporation of disk material into planets. 
Planet-disk interactions may be very important for setting planetary compositions; for example, pressure bumps generated by young planets could cause a pile-up of volatile-rich grains from the outer disk that are then prevented from reaching inner rocky planets.  
%While observation of planet-disk interactions have been observed for a handful of the brightest disks (mostly around Herbig Ae/Be stars), 
While the outer regions of bright disks have been imaged in the optical to near-IR,
ELTs will enable us to probe structures in fainter disks and structures much closer to the star where most planets appear to reside (Fig.~\ref{fig:substructure}). Direct observations of disk-planet interactions are needed to tease out the essential physics behind planet formation.

\begin{figure}[bt]
    \centering
    \includegraphics[width=\textwidth,trim=0.15in 3.1in 0.15in 3.1in,clip=true]{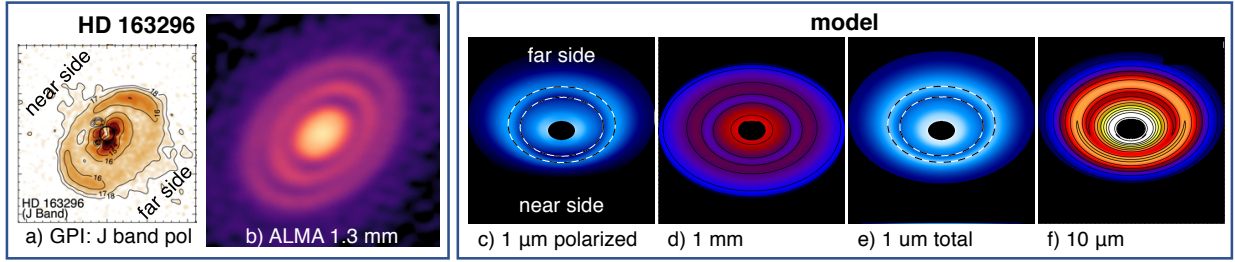}
\vspace*{-5ex}    
\caption{\label{fig:diskimages} \em
Observed structure in HD 163296 compared with a disk model with a 70 Earth mass planet at 10 au inclined at 45$^{\circ}$. 
The (a) polarized J-band image taken with GPI (Monnier et al.~2017)
and the (b) ALMA image at 1.3 mm (Isella et al.~2016)
can be compared to the model disk as viewed in (c) polarized intensity at 1 $\mu$m \citep{JangCondell2017} and at (d) 1 mm \citep{JangCondellTurner2013}, respectively.  The images on the right show the disk model viewed at (e) 1 and (f) 10 micron total intensity \citep{JangCondellTurner2013}.
    An ELT operating in the mid-infrared could probe thermal emission from the disk surface, as shown in the 10 micron image.
    }
    \vspace*{-2ex}
\end{figure}

Resolved imaging of young, gas-dominated protoplanetary disks reveal structures such as gaps (e.g., HL Tau [Fig.~\ref{fig:substructure}], HD 163296 [Fig.~\ref{fig:diskimages}]), spiral arms (e.g., SAO 206462 [Fig.~\ref{fig:substructure}]), warps \citep{Loomis2017,Mayama2018,Benisty2018}, and inner holes \citep{vandermarel2015}.
Hydrodynamic simulations predict that gaps and spiral arms can be caused by planets embedded in and interacting with the disk  \citep{Dong2015gaps,Dong2015spiralplanet}. The appearance of gaps and spiral arms can also reveal planet properties \citep{Debes2013,Dong2015spiralplanet} or indicate gravitational instability \citep{DongNajitaBrittain18}.
Other simulations have shown that a close-in stellar or planetary companion can break up the inner and outer disk and excite a misaligned inner disk, or warp \citep{Facchini:2013, Nealon:2018}.
All these large scale disk structures are illuminated by the central star and may also cast shadows in disks: features that can be seen in resolved imaging with ELTs either in scattered light in optical to near-IR or in thermal emission at $10-20$ $\mu$m (see Fig.~\ref{fig:diskimages}).
For example, shadows cast by warps could produce apparent spiral arms in the outer disk \citep{Benisty:2017, Benisty:2018, Min:2017}. 

%both in ALMA observations, directly by scattered light, and indirectly by shadowing \citep{Marino2015,Pinilla2018}.  

%Some protoplanetary disks are optically thin in their inner regions, suggesting that the disks are in the process of being cleared from the inside-out, and for this reason they are called transitional disks. Planets could also be an explanation for the clearing of these inner holes.

% Imaging and spectroscopic observations at mm-wavelengths with ALMA have revealed rings and gas motions consistent with planets \citep[e.g.][]{Tang:2017,Teague:2018}.
Multiwavelength imaging can be particularly useful for interpreting disk structure. 
%Gap sizes and depths can constrain the properties of disks and embedded planets, particularly if imaged at multiple wavelengths \citep{JangCondellTurner2012}. 
For example, a gap at 50 au has been imaged in HD 163296 in both polarized scattered light and 1.3 mm continuum emission (Fig.~\ref{fig:diskimages}).
By combining these datasets, the near side/far side degeneracy can resolved and the disk scale height estimated.  
As ALMA continues to reveal more gaps, rings, and spirals in disks \citep[e.g.][]{DSHARP2018}, synergy with imaging from ELTs at similar angular scales can give us greater insight into planets forming in them.

\vspace{-2ex}
\section{Survey Strategy}
\vspace{-2ex}
We recommend an unbiased imaging and spectroscopy survey of disk-bearing (Class~II) stars  in the nearby young (1--10~Myr) star-forming regions of    Taurus, Ophiuchus, Lupus, and Scorpius-Centaurus.  
% Studies to date have been limited to the brightest, most extreme (e.g., in mass), most photogenic systems, and ELTs provide the opportunity to overcome this limitation.  
Likewise, the ability to combine both spectroscopic dissection of the physical and chemical properties of disks along with high-contrast imaging of the same disks and their protoplanets will provide an unbiased, holistic view of planet formation as it occurs.

Low spectral resolution Spitzer-IRS observations demonstrate that molecular detections will be nearly ubiquitous for young disks with sufficient S/N \citep{Pontoppidan2010}.  The addition of high spectral and spatial resolution will be crucial for disentangling modeling degeneracies and determining inner disk chemical abundance structures.  Most of these molecules have been detected at high spectral resolution in only a handful of the brightest disks \citep[e.g.][]{Mandell2012,Gibb2013,Najita2018}. ELTs will enable the observation of a more representative sample of disks, revealing the more ``typical'' conditions of planet formation.

%An an example of the rich possibilities afforded by the ELTs, we consider what it would take for a survey in Ophiuchus, one of the nearest sites of ongoing star formation age ages $\lesssim$3 Myr. There are currently known $\sim$300 young stars in that cluster \citep{Wilking:2008}. 
%The lowest mass members are probably amongst those that do not even have spectral typing due to high extinction and veiling. 
%Infrared spectroscopy has the advantage of penetrating both extinction and looking at wavelength regions less affected by veiling. All the known stars have K $<$ 15 because they were identified, in part, through the 2MASS catalog. The predictions for laser guide star AO on the ELTs is that it will work with any star of K$<$18 mag using the target itself as the tip-tilt star.

The sensitivity required is a function of both the continuum brightness and the need to detect small line:continuum ratios. As an example, the median Taurus Class II star has J=9.5~mag and W1 (3.4~$\mu$m)=7.2~mag.  We can scale this down to stars with M$<$0.5 M$_\odot$, which are the most interesting targets because because of their high fraction of low-mass planets, but the least explored by the current generation of telescopes at high spectral resolution (e.g., only two low-mass stars were studied in the sample of 55 stars in \citet{Banzatti:2017}). The low mass stars in Taurus have J$\sim$14.8~mag. We can use J-[W1] to estimate the disk brightness, or, more realistically, the fact that disk mass is roughly proportional to stellar mass \citep{Andrews:2013}, and assume the disks around low mass stars
will be somewhat fainter, with W1 $\sim$ 13 mag.  This sensitivity can be achieved with GMTNIRS and TMT-MODHIS at S/N$\sim$50 in 1 hr. This same sensitivity can detect a line:continuum ratio of 0.06, comparable to what is achieved presently on bright targets with current generation telescopes.

With these spectra we will be able to compare the chemistry of disks relative to system parameters such as stellar spectral type, stellar accretion rate, disk geometry, and x-ray luminosity. One of the principal outcomes of this survey will be the testing of thermochemical models of disks (Fig. \ref{fig:chem_abund}). 
In addition to using the spectra to study the chemistry of disks, the line profiles will be analyzed spectro-astrometrically (or possibly with an IFU) to identify signatures of disk winds (e.g. \citealt{Pontoppidan2011}), planet-disk interactions, and possibly emission from a circumplanetary disk itself (e.g., \citealt{Brittain:2014}). The indirect signatures of disks inferred from the spectra will then be validated by direct imagery of forming planets.

%{\bf Spectroscopy:} We will acquire velocity and spatially resolved spectra from 1-5~$\mu$m using GMTNIRS and 3-13~$\mu$m using MICHI. These regions include ro-vibrational and rotational transitions from the main molecular carriers in disks: CO, CO$_2$, H$_2$O, OH, HCN, C$_2$H$_2$, NH$_3$, and CH$_4$ (\citealt{Najita2017}; see Figures \ref{fig:spectrum} and \ref{fig:chem_abund})\footnote{H$_2$ is more abundant, but can only easily be observed in the vacuum UV from space, see, e.g., \citet{Bitner2007,France2013}}. 

%{\bf Imaging:}  Deep IR imaging of our sample will enable direct detection of
%protoplanets, with follow-up characterization using multi-band
%photometry to measure their SEDs and IR spectroscopy to measure physical
%properties (temperatures, surface gravities, composition, and
%accretion).  ELTs are potentially sensitive to protoplanets of
%$\sim$1~Jupiter mass within 30~AU, even for extincted objects still
%embedded in the disk material. Around the nearest stars, ELTs can probe
%1--10~AU separations, which are expected to contain the bulk of forming
%gas giants, and all targets will be well-imaged at 10--100's of AU where
%ALMA observations of disk substructure suggest massive planets have
%formed.

% Let's assume we want to detect 2Myr Jupiters - what would be a reasonable inner working angle+contrast? Can we say something interms of M\.{M}? 

% Disks are the most challenging targets for ground-based AO-assisted imaging because uncorrected atmospheric distortions create halos of light around the stars, in just the locations where the disks are to be found. 

Disks with known well-resolved substructure from ALMA will naturally be
higher priority targets for high-contrast imaging (e.g.,
Figure~\ref{fig:diskimages}).
%both to search for protoplanets and
%obtained detailed multi-wavelength imaging of disk substructures.  
While
there are a few dozen such ALMA-resolved disks known now
\citep[e.g.\/][]{alma2015,Andrews:2016,DSHARP2018}, we can
anticipate many more by the time the ELTs are conducting science
observations.
% One way to characterize the brightness of disks in scattered light is their surface brightnesses (at some distance) divided by the total stellar brightness. % For example, the disks imaged with SPHERE and GPI have inte
The challenge for the ELTs will be to find structures, such as spirals, rings, and warps, in the inner regions of disks, where the strongest disk-planet interactions are expected and where they have the most impact on the composition and architectures of planetary systems. At the distance to the nearest clusters of on-going star formation, at $\sim$140 pc, the diffraction-limited resolution of 30~m class telescopes is essential to resolving the important scales: resolving 1 AU at 140 pc requires an angular resolution of 7 mas, which is the diffraction limit of a 30~m telescope at a wavelength of 1 $\mu$m.  An inner working angle of 2-3 $\lambda/D$, provides a view from just inside to outside the putative water ice line. At visible and near-infrared wavelengths, photons are efficiently scattered off of dust in the disk surface and in imaging observations can reveal warps and spirals in the dust distribution. Using spectroastrometry, the gas motions in the vicinity of such structures can be measured and indicate whether a planet is truly modifying the trajectories of disk material.
%Emission arising from an eccentric orbit results in a non-varying asymmetric velocity profile \citep[e.g.][]{Regaly:2010}. This signature has been observed in HD~100546 \citep{Liskowsky:2012,Brittain:2014} and MWC 758 \citep{Dong:2018}. 
In one hour, the point source 
sensitivity for coronagraphic imaging is 15.6 magnitudes at 
L-band, or about 0.05 Jy/asec$^2$.  
As shown in Fig.~\ref{fig:diskimages}, a planet at 10 au around a 
solar mass star
can create structures that are as bright as $\sim0.1$ Jy/asec$^2$, 
which is easily seen with this sensitivity. 
%Structures due to planet-interactions are most easily found in face-on
%disks, which means that standard AO observing technique of angular
%differential imaging is challenging. Because disks polarize light when
%they scatter, differential polarized imaging does work efficiently to
%distinguish disk scattering from residual stellar speckles left %uncorrected by the adaptive optics system, which is
%why both SPHERE and GPI have produced stunning disk images with their
%polarimetric modes.

\pagebreak
%\textbf{References}

\bibliographystyle{aasjournal}
\bibliography{references.bib}

\end{document}